\documentclass[aps,pra,twocolumn,footinbib,superscriptaddress]{revtex4-1}

\usepackage[colorlinks=true,bookmarks=false,citecolor=blue,urlcolor=blue]{hyperref}

\usepackage{amsmath,amssymb,esint}
\usepackage{graphicx}
\usepackage{verbatim}
\usepackage{color}
\usepackage[english]{babel}
\usepackage{bm}

\usepackage{epstopdf}

\def\dfrac{\displaystyle\frac}
\newcommand{\eps}{\varepsilon}
\renewcommand {\phi}{\varphi}

\newcommand*\VF[1]{\mathbf{#1}}
\newcommand*\dif{\mathop{}\!\mathrm{d}}

\begin{document}
\title{Multipolar second-harmonic generation by Mie-resonant dielectric nanoparticles} % Second-harmonic generation by optically resonant high-index dielectric nanoparticles

\author{Daria Smirnova}
\affiliation{Institute of Applied Physics, Nizhny Novgorod 603950, Russia}
\affiliation{Nonlinear Physics Centre, Australian National University, Canberra ACT 2601, Australia}

\author{Alexander I. Smirnov}
\affiliation{Institute of Applied Physics, Nizhny Novgorod 603950, Russia}

\author{Yuri S. Kivshar}
\affiliation{Nonlinear Physics Centre, Australian National University, Canberra ACT 2601, Australia}

\begin{abstract}

By combining analytical and numerical approaches, we study resonantly enhanced second-harmonic generation (SHG) by individual high-index dielectric nanoparticles made of centrosymmetric materials. Considering both bulk and surface nonlinearities, we describe second-harmonic nonlinear scattering from a silicon nanoparticle optically excited in the vicinity of the magnetic and electric dipolar resonances. We discuss the contributions of different nonlinear sources, and the effect of the low-order optical Mie modes on the characteristics of the generated far-field. We demonstrate that the multipolar expansion of the radiated field is dominated by {\em dipolar} and {\em quadrupolar} modes (two axially symmetric electric quadrupoles, an electric dipole, and a magnetic quadrupole), and the interference of these modes can ensure directivity of the nonlinear scattering. The developed multipolar analysis can be instructive for interpreting the far-field measurements of the nonlinear scattering, and it provides prospective insights into a design of CMOS-compatible nonlinear nanoantennas fully integrated with silicon-based photonic circuits, as well as new methods of nonlinear diagnostics.
\end{abstract}

\keywords{nanophotonics, Mie resonances, second-harmonic generation, nonlinear scattering, silicon nanoparticle}

\pacs{42.65.−k, 78.35.+c, 42.70.Nq} 

\maketitle

\section{Introduction}

Being stimulated by a rapid progress in nanofabrication techniques, dielectric resonant nanostructures with high refractive index are currently employed in various applications of nanophotonics, offering competitive alternatives to plasmonic nanoparticles~\cite{kuznetsov2016science}. Advantageous optical properties of high-index dielectric nanoparticles, such as
low dissipative losses, optical magnetic response, and multipolar resonances,
imply exclusive capabilities for light manipulation at subwavelength scales,
especially in the nonlinear regime~\cite{Smirnova2016}.

Acting as optical nanoantennas, high-permittivity dielectric nanoparticles exhibit strong interaction with light due to the excitation of both electric and magnetic Mie resonances they support. Compared to plasmonic nanoscale structures, where the electric field is strongly confined to surfaces, the electric field of the resonant modes in dielectric nanoparticles penetrates deep inside their volume, thus enhancing intra-cavity light-matter interactions in a bulk material. Such a strategy of utilizing the Mie resonances in the subwavelength dielectric geometries has been recently recognized as a promising route for improving the nonlinear conversion processes
at the nanoscale~\cite{Shcherbakov:2014:NL,Yang2015,Smirnova:2016:ACS-Ph,Shorokhov2016,LiuBrener2016,Grinblat:2016:NL,CamachoMorales2016} .

Second-harmonic generation in plasmonic nanostructures is known to be governed mainly by surface nonlinear response, which can be enhanced at the geometric plasmon resonances~\cite{Dadap1999,Dadap2004,Gonella2011,Thyagarajan2012,Butet2012,Kauranen2012,Capretti2013,Biris2013,Smirnova2014,Butet2014,Butet2015}. Primarily, electric dipole response associated with the surface plasmon resonance is most
widely exploited for deeply subwavelength metallic particles and their composites, and the
nonlocal bulk contribution to second-harmonic generation (SHG) is largely ignored~\cite{Wang2009,Bachelier2010}.
The excitation of multipolar resonances driven by displacement currents in dielectric nanostructures can significantly reshape the nonlinear scattering, in particular, due to the bulk nonlinear response altered with the field gradients distributed over the volume.
One of the most promising material for implementation of all-dielectric nanophotonics is silicon due to its CMOS compatibility and strong optical nonlinearities~\cite{leuthold2010nonlinear,priolo2014silicon}. In particular, silicon was employed in most of works on the trapped magnetic dipole resonances~\cite{Evlyukhin2012,Kuznetsov2012} and the associated enhancement of the third-order nonlinear processes~\cite{Shcherbakov:2014:NL,Shcherbakov2015_2,Yang2015,Shorokhov2016,Wang2017}.
Though silicon, both crystalline and amorphous, is a centrosymmetric material, and thus, similar to noble metals, its bulk second-order nonlinear response is inhibited~\cite{Cazzanelli2016}, the light confinement and enhancement due to excitation of the resonant modes
increases the efficiency of the frequency conversion, and quite high yield of SHG from individual nanowires~\cite{Wiecha2015,Wiecha2016} and nanoparticles~\cite{Makarov2017} can be achieved.

\begin{figure} [b]
\includegraphics[width=0.95\columnwidth]{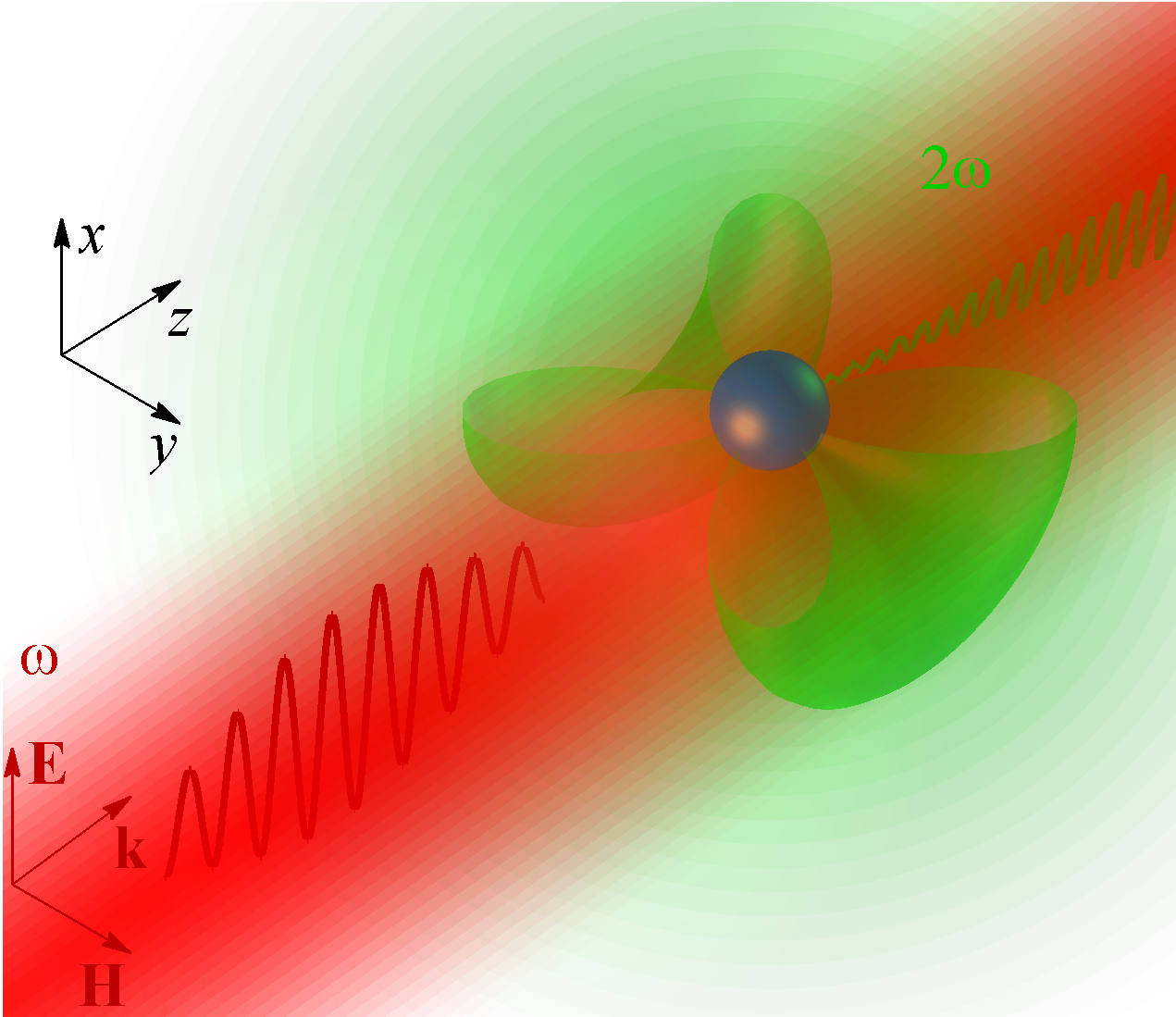} % ModelVisualizationCropped2.png
\caption{ (Color online) Schematics of the geometry. Laser radiation of frequency $\omega$ is scattered by the nanoparticle. As a result of nonlinear
interaction, the second-harmonic light of $2\omega$ is generated.}   \label{fig:fig1}
\end{figure}

In this paper, we investigate the characteristic features of SHG from dielectric nanoparticles made of high-index centrosymmetric materials and optically excited in the vicinity of the pronounced low-order Mie resonances, with a particular focus on the magnetic dipole resonance. We take into account the contributions of both surface and bulk induced nonlinear sources described in the framework of the phenomenological model~\cite{Mochn2003,Wiecha2015,Wiecha2016}. We reveal that the SH radiation is dominated by dipolar and quadrupolar contributions, specifically by two axially symmetric electric quadrupoles (oriented along the magnetic and electric fields, respectively, in the incident wave), an electric dipole (directed along the wave vector of the incident wave), and a magnetic quadrupole. We emphasize that the case we study is essentially distinct from the Rayleigh limit, small plasmonic particles and Rayleigh-Gans-Debye model, or first Born approximation (assuming a low refractive-index mismatch between the interior of the particle and a host medium)~\cite{Dadap2004,Shan2006,Wunderlich2011}. By contrast, in the small-particle limit the SH field is described by one electric quadrupole and one electric dipole~\cite{Dadap2004}. In the experimental study~\cite{Shan2006}, it is further discussed that the octupolar traits in SH scattering diagrams appear for the nonresonant polysterene nanoparticle %of radius $a=85$~nm 
as corrections compared to the Rayleigh-limit SHG~\cite{Dadap2004} % (relying on the small ratio $k(\omega)a$) observed from smaller nanoparticles of radius $a=50$~nm 
as a consequence of increasing the size parameter.

Here we derive the excitation coefficients of the nonlinearly generated multipoles with an original procedure based on the use of the Lorentz lemma. It can be regarded as a more practical alternative to the nonlinear Mie theory analysis~\cite{Dadap2004,deBeer2009,Gonella2011,Capretti2013}. Our approach can be applied to Mie-resonant nanoparticles made of not only centrosymmetric but also noncentrosymmetric high-index materials actively employed for nonlinear nanophotonics~\cite{CamachoMorales2016,Kruk2017,Timpu2017,Ma2017}. In addition, we provide a detailed analytical solution for the resonantly enhanced SHG driven by the pronounced magnetic dipole excitation with the approach outlined in Ref.~\cite{Smirnova:2016:ACS-Ph}. The validity of the developed theory and analytically described multipolar expansion of the SH field is confirmed in the direct full-wave numerical calculations.
% Despite approximations made, the developed theory agrees well with the finite-element-method numerical modeling, 

\section{Multipolar analysis of nonlinear scattering}

\begin{figure} [t]
\includegraphics[width=0.95\columnwidth]{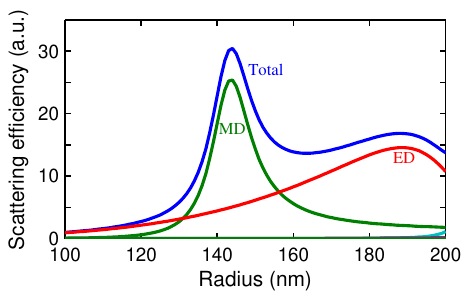} % scattering_cross-section_at_ff.png
\caption{ (Color online)
Linear scattering efficiency (blue curve) calculated for a spherical silicon nanoparticle at the pump wavelength $\lambda_0=1050$~nm. The labeled MD and ED
contributions are shown by green and red lines.}   \label{fig:linscat}
\end{figure}

We consider a high-permittivity spherical dielectric particle of radius $a$, excited by the linearly-polarized plane wave ${\bf E} ({\bf r}) = \hat{{\bf x}} E_0 e^{ik_0z}$  propagating in the $z$ direction, as illustrated schematically in Fig.~\ref{fig:fig1}. The analysis we perform also gives a qualitatively correct picture of the second-harmonic (SH) fields generated by an arbitrary single-scale nanoscale object (e.g., a finite-extent nanorod whose cross-sectional diameter is of the order of its length). The particle is characterized by the frequency-dependent dielectric constant $\eps(\omega)$. The homogeneous host medium is air. The problem of linear light scattering by a sphere is solved using the multipole expansion in accord with Mie theory. The resultant scattering efficiency is plotted in Fig.~\ref{fig:linscat} for a silicon nanoparticle excited at wavelength $\lambda_0=1050$~nm in the range of radii featuring MD and ED resonances.

In the frequency range between the magnetic (MD) and electric (ED) dipolar resonances, the electric field at the fundamental frequency inside the nanoparticle is well
approximated by a superposition of only MD and ED modes, as evidenced by Fig.~\ref{fig:linscat}
% \begin{widetext}
{\small{
\begin{multline} \label{eq:eq1}
{\bf{E}}_{\text{in}}^{\left( {\omega} \right) }\! \approx \!  {E_0}  \left[ \dfrac{1}{k(\omega)} A_{1,1}^E \nabla \times j_1{(k(\omega)r)} \left( {\bf{X}}_{1,1}( {\theta ,\phi }) - {\bf{X}}_{1,-1}( {\theta ,\phi }) \right) \right. \\ \left. + A_{1,1}^M  j_1{(k(\omega)r)}  \left( {\bf{X}}_{1,1}( {\theta ,\phi }) + {\bf{X}}_{1,-1}( {\theta ,\phi }) \right)  \right] \:,
\end{multline}
}}
% \end{widetext}
where $k(\omega)=k_0\sqrt{\eps(\omega)}$ is wavenumber in the medium, $k_0 = \omega/c$,
$j_1 (k(\omega)r)$ is spherical Bessel function of order $l=1$,  $ {\bf{X}}_{1,1}( {\theta ,\phi })$
%  $ {\bf{X}}_{1,1}( {\theta ,\phi }) = \dfrac{1}{4} \sqrt{\dfrac{3}{\pi}} \{ 0, 1, i \cos{\theta} \} \text{exp}(i \phi)$   is the vector spherical harmonic of degree $l=1$ and order $m=1$,
are vector spherical harmonics (in the spherical coordinate system associated with $z$ axis),
$A_{1,1}^E$  and $A_{1,1}^M$ are coefficients known from Mie theory~\cite{Jackson1965}.

The pronounced character of the low-order Mie resonances is essential for many applications of high-permittivity dielectric nanoparticles in low-index environment~\cite{kuznetsov2016science,Kruk2017_2} and for the analysis we develop below. We specifically focus on Mie-resonant dielectric nanoparticles, whose sizes correspond to the resonant excitation of the leading magnetic dipole and electric dipole modes at the laser fundamental wavelength, as shown in Fig.~\ref{fig:linscat}. The analysis of SHG from high-index dielectric nanoparticles exhibiting dipolar resonances is important for modern nanoscale optics, given the increasing interest in the rapidly expanding field of all-dielectric nanophotonics and growing number of nonlinear experiments being currently done by many research groups worldwide exactly under the conditions associated with resonant excitation of the low-order Mie modes~\cite{Smirnova2016}.

The second-order polarization for the particles made of centrosymmetric homogeneous materials %(e.g., amorphous silicon)
can be written as a superposition of dipolar surface (local) and quadrupolar bulk (nonlocal) contributions~\cite{guyot1986general,guyot1988bulk, Mochn2003,Wiecha2015,Wiecha2016}
\begin{subequations} \label{eq:PNL}
\begin{align}
{\bf{P}}^{\left( {2\omega } \right)}  = {}& {\bf{P}}_{\text{surf}}^{\left( {2\omega } \right)} +  {\bf{P}}_{\text{bulk}}^{\left( {2\omega } \right)} \text{ , }\\
\begin{split}
{\bf{P}}_{\text{surf}}^{\left( {2 \omega } \right)} = {}&
% \varepsilon_0
\delta(r-a+0) \left[ {\bf{\hat{r}}_0} ( \chi^{\left( {2} \right)}_{ \bot  \bot  \bot } (E_{r}^{( \omega )})^2  +    \chi^{\left( {2} \right)}_{\bot \parallel \parallel}  (E_{\tau}^{( \omega )})^2)   \right. \\
 & \left. + 2 {\bm{\hat{\tau}}_0}  \chi^{\left( {2} \right)}_{\parallel  \bot \parallel } E_{r}^{( \omega )}{E}_{\tau}^{( \omega )} \right],
 \end{split} \label{eq:PNLb} \\
 \begin{split}
{\bf{P}}_{\text{bulk}}^{\left( {2\omega } \right)} = {}&
% \varepsilon_0
\left[ \beta {\bf{E}}^{( \omega )} \nabla \cdot {\bf{E}}^{( \omega )} +
   \gamma \nabla \left( {{{\bf{E}}^{\left( \omega  \right)}} \cdot {{\bf{E}}^{\left( \omega  \right)}}} \right) \right. \\
 & \left. + \delta' \left( {{{\bf{E}}^{\left( \omega  \right)}} \cdot \nabla } \right){{\bf{E}}^{\left( \omega  \right)}} \right] \Pi (a-0-r)\:,
 \end{split} \label{eq:PNLc}
\end{align}
\end{subequations}
% Here, $\chi^{(2)}$, $\beta$, $\gamma$ and $\delta'$ are the material parameters characterizing optical nonlinearity, ${\bf{r}}_s$ defines the interface surface, and $\varepsilon_0$ is the dielectric permittivity of vacuum. Electric field vector \textbf{E}$^{( \omega )}$ at the fundamental wavelength can be decomposed near the interfaces onto two unit vectors ${\bf{\hat{n}}}$ and ${\bm{\hat{\tau}}}$ with corresponding normal \textit{E}$_n^{( \omega )}$ and tangential \textit{E}$_{\tau}^{( \omega )}$ components relatively to the surface. Vanishing for bulk homogeneous media, the $\beta$ term leads to an additional quadrupolar surface polarization because of discontinuity of the normal electric field component near the interfaces~\cite{Kulyuk1991}.
% The $\gamma$ term also exhibits a surface-like behavior, and it is often referred to as a nonseparable bulk contribution~\cite{Wiecha2016}. Overall, due to a composite nature, the polycristalline Si nanoparticle acquires an effective SH source distributed over its volume, that comes from the average nonlinear response of nanograins with account for their interfacial nonlinear polarizations~\cite{Mochn2003}.
where $E_{r}^{( \omega )}$ and $E_{\tau}^{( \omega )}$ are the radial and tangential components
of the electric field on the spherical surface, $\bf{\hat{r}}_0$, $\bm{\hat{\tau}}_0$ are the corresponding unit vectors. The coefficients $\gamma$, $\beta$, $\delta'$, $\chi^{\left( {2} \right)}_{ \bot  \bot  \bot }$, $\chi^{\left( {2} \right)}_{\bot \parallel \parallel} $, $\chi^{\left( {2} \right)}_{\parallel  \bot \parallel }$ are material parameters of the dielectric, the $\beta$ term vanishes in the bulk, $\nabla \cdot {\bf{E}}^{( \omega )} =0$, due to the homogeneity of the material, $\delta(\tilde{r})$ is the Dirac delta function, and step function $\Pi(\tilde{r})$ is defined by $\Pi(\tilde{r}) = \left\{ 0, \tilde{r}<0 ; 1, \tilde{r} > 0 \right\}$. Importantly, the $\gamma$ term exhibits a surface-like behavior, and it is often referred to as a nonseparable bulk contribution~\cite{Capretti2013, Wiecha2016}. We assume phenomenological model~\eqref{eq:PNL}  qualitatively valid for amorphous and crystalline silicon nanoparticles, disregarding any anisotropy effects~\cite{Mochn2003,Wiecha2015,Wiecha2016}.
Specifics of SHG from nanocrystalline silicon nanoparticles was studied experimentally and numerically in Ref.~\cite{Makarov2017}.
According to Eq.~\eqref{eq:PNLb}, the nonlinear surface sources ${\bf{P}}_{\text{surf}}^{\left( {2 \omega } \right)}$ are defined by the field ${\bf{E}}^{( \omega )} $  at the pump wavelength inside the nanoparticle. Introducing functions $\delta(r-a+0)$ and $\Pi (a-0-r)$ in Eqs.~\eqref{eq:PNLb} and~\eqref{eq:PNLc} allows us to formalize mathematical derivations.

Plugging Eqs.~\eqref{eq:PNL} into the Maxwell's equations,
the SH electromagnetic field $ { {\bf{E}}^{\left( 2\omega  \right)} }, {\bf{H}}^{\left( 2\omega  \right)}$ is the forced solution of a set of equations
\begin{subequations}  \label{eq:3}
\begin{align}
\nabla \times {\bf E}^{(2\omega)}  & = 2i k_0 {\bf H}^{(2\omega)}, \label{eq:3a} \\
\nabla \times {\bf H}^{(2\omega)}  & = -2i k_0 \eps^{(2\omega)} (r) {\bf E}^{(2\omega)} + \dfrac{4\pi}{c} {\bf j}^{({2\omega})} \:, \label{eq:3b}
\end{align}
\end{subequations}
where $ {\bf j}^{({2\omega})} = -2i\omega ( {\bf{P}}_{\text{surf}}^{\left( {2\omega } \right)} +  {\bf{P}}_{\text{bulk}}^{\left( {2\omega } \right)} )$ is the current density induced due to the quadratic nonlinearity,
and
\begin{equation}
\eps^{(2\omega)} (r) =
\begin{cases}
\eps (2\omega), & r \le a\\
1, & r>a\,
\end{cases}
\end{equation}
is the dielectric permittivity distribution at the second harmonic frequency. Note, in the considered frequency range under approximation~\eqref{eq:eq1}
the polarization sources, and consequently, the external current ${\bf j}^{({2\omega})}$ constitutes the quadratic form of the electric field ${\bf{E}}^{\left( {\omega} \right) }$, which is defined predominantly by the electric and magnetic dipolar modes excited at the fundamental frequency $\omega$. Since these two modes depend linearly on sine and cosine functions of of the polar angle, multipolar expansion of the generated SH field to the leading order contains only dipolar and quadrupolar spherical harmonics.

\begin{figure} % [t]
\includegraphics[width=0.97\columnwidth]{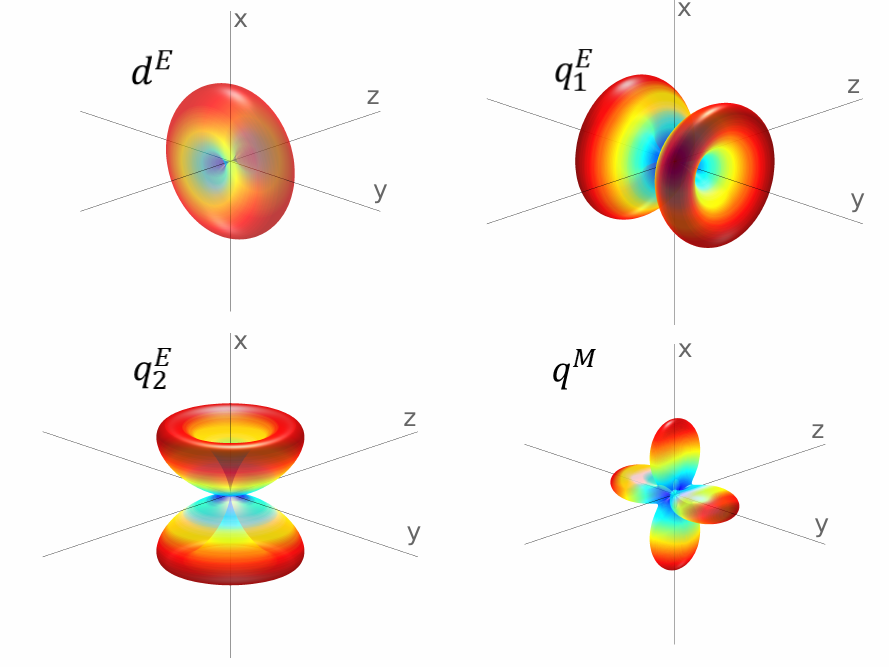} % image096.jpg
\caption{(Color online) Radiation patterns of the nonlinearly generated SH multipoles: $z$-aligned electric dipole $d^E$, $y$-axially symmetric electric quadrupole $q^E_1$, $x$-axially symmetric electric quadrupole $q^E_2$, and magnetic quadrupole $q^M$.
}  \label{fig:figRPs}
\end{figure}

Similar to the work~\cite{Kruk2017}, we analyze the induced nonlinear multipolar sources by
employing general expressions for the electric and magnetic multipolar coefficients at the SH wavelength as defined by the overlap integrals of the sources with spherical harmonics~\cite{Jackson1965}. Our calculations show that within the framework of approximation~\eqref{eq:eq1}, the multipolar composition features two axially-symmetric electric quadrupolar (EQ)
components, whose amplitudes are proportional to  $(A_{1,1}^E )^2$  and  $(A_{1,1}^M )^2$, as well as ED and MQ modes with amplitudes proportional to $(A_{1,1}^E A_{1,1}^M)$.
Thus, outside the nanoparticle, the SH magnetic field assumes the form
\begin{widetext}
\begin{multline} \label{eq:SHMField}
{\bf{H}}^{\left( {2\omega} \right) }(r>a)\approx  {E_0}^2  \left[(A_{1,1}^M )^2 q^E_1 h^{(1)}_2 (2k_0 r) {\bf{X}}_{2,0} (\theta_1) + (A_{1,1}^E )^2 q^E_2 h^{(1)}_2 (2k_0 r) {\bf{X}}_{2,0} (\theta_2) + \right. \\ \left. A_{1,1}^E A_{1,1}^M d^E  h^{(1)}_2 (2k_0 r) {\bf{X}}_{1,0} (\theta) - \dfrac{i}{2k_0} A_{1,1}^E A_{1,1}^M q^M \nabla \times \left( h^{(1)}_2 (2k_0 r)  \left\{  {\bf{X}}_{2,1} (\theta_1,\phi_1) - {\bf{X}}_{2,-1} (\theta_1,\phi_1)  \right\} \right) \right]\:.
\end{multline}
\end{widetext}
Here $ {\bf{X}}_{l,m}$ are spherical functions; $( {\theta ,\phi })$, $( {\theta_1 ,\phi_1})$, $( {\theta_2 ,\phi_2})$ are polar and azimuthal angles of the spherical coordinate systems associated with $z$, $ y$ and $x$ axes, respectively, $h^{(1)}_l$ is the spherical Hankel function of the first kind of order $l$. In Eq.~\eqref{eq:SHMField}, the terms proportional to $q^E_1$ and $q^E_2$ describe the fields emitted by the electric quadrupoles which are axially symmetric to the $y$ and $x$ axes, the term proportional to $d^E$ is the radiation field of the electric dipole oriented along the propagation direction $z$ of the incident wave, and the term proportional to $q^M$ is due to the presence of the magnetic quadrupolar component in the source. The far-field diagrams of the generated SH multipoles,
\begin{equation*}
\begin{aligned}
&F_{d^E} (\theta)  \propto \sin^2{\theta}\:,  \\
&F_{q^E_1} (\theta_1) \propto  \sin^2{(2\theta_1)} \:,  \\
&F_{q^E_2} (\theta_2) \propto   \sin^2{(2\theta_2)} \:, \\
&F_{q^M}(\theta_1,\phi_1)  \propto \cos^2{(2\theta_1)}\sin^2{\phi_1} + \cos^2{\theta_1}\cos^2{\phi_1}  \:,
\end{aligned}
\end{equation*}
are visualized in Fig.~\ref{fig:figRPs}.

The excitation coefficients of the multipolar modes, $q^E_1$, $q^E_2$, $d^E$, $q^M$, are linear functions of the phenomenological parameters,  $\gamma$, $\beta$, $\delta'$, $\chi^{\left( {2} \right)}_{ \bot  \bot  \bot }$, $\chi^{\left( {2} \right)}_{\bot \parallel \parallel} $, $\chi^{\left( {2} \right)}_{\parallel  \bot \parallel }$.
Analytical expressions for the multipolar amplitudes can be found using the Lorentz lemma~\cite{Vainshtein,Jackson1965}. The Lorentz lemma is widely applied in electrodynamics for calculation of amplitude coefficients of the guided modes excited by external sources and radiation diagrams of emitters. Here we show that the methodology based on the Lorentz lemma can be adopted for the analysis of the nonlinear scattering. This approach facilitates mathematical derivations, especially, in the treatment of the bulk nonlinearity, and, more importantly, it allows for generalization to nanoparticles of nonspherical shapes. For our problem, it can be formulated as follows. We introduce the auxiliary electromagnetic field $ \left\{ {\bf{E}}^{\left( {2 \omega} \right) }_1,{\bf{H}}^{\left( {2 \omega} \right)}_1 \right\}$ satisfying the Maxwell's equations in the medium with the dielectric permittivity $\eps^{(2\omega)} (r)$ in the absence of the external sources
\begin{subequations}  \label{eq:6}
\begin{align}
\nabla \times {\bf E}^{(2\omega)}_1 & = 2i k_0 {\bf H}^{(2\omega)}_1, \label{eq:6a} \\
\nabla \times {\bf H}^{(2\omega)}_1  & = -2i k_0 \eps^{(2\omega)} (r) {\bf E}^{(2\omega)}_1 \:.\label{eq:6b}
\end{align}
\end{subequations}
We then apply scalar multiplication to Eqs.~\eqref{eq:3a} and~\eqref{eq:6b} by ${\bf H}^{(2\omega)}_1$ and ${\bf E}^{(2\omega)}$, respectively, and subtract one from another
\begin{multline} \label{eq:LL1}
{\bf H}^{(2\omega)}_1  \nabla \times  {\bf E}^{(2\omega)}  - {\bf E}^{(2\omega)} \nabla \times {\bf H}^{(2\omega)}_1 \\ =2 i k_0  \left( \eps^{(2\omega)} (r) {\bf E}^{(2\omega)}_1 {\bf E}^{(2\omega)} + {\bf H}^{(2\omega)}_1 {\bf H}^{(2\omega)} \right)  \:.
\end{multline}
In a similar manner, we find
\begin{multline} \label{eq:LL2}
{\bf H}^{(2\omega)} \nabla \times {\bf E}^{(2\omega)}_1   - {\bf E}^{(2\omega)}_1  \nabla \times {\bf H}^{(2\omega)} \\ = 2 i k_0 \left( \eps^{(2\omega)} (r) {\bf E}^{(2\omega)}_1 {\bf E}^{(2\omega)} +  {\bf H}^{(2\omega)}_1 {\bf H}^{(2\omega)} \right) \\- \dfrac {4\pi}{c} {\bf j}^{({2\omega})} {\bf E}^{({2\omega})}_1\:.
\end{multline}
Subtracting Eq.~\eqref{eq:LL2} from Eq.~\eqref{eq:LL1}, we obtain
\begin{multline} \label{eq:LL3}
\nabla \cdot [ {\bf E}^{(2\omega)} \times {\bf H}^{(2\omega)}_1 ] - \nabla \cdot [ {\bf E}^{(2\omega)}_1 \times {\bf H}^{(2\omega)} ] = \dfrac {4\pi}{c} {\bf j}^{({2\omega})} {\bf E}^{({2\omega})}_1  \:,
\end{multline}
as a consequence of the Lorentz lemma. We next integrate Eq.~\eqref{eq:LL3} over the volume $V$, bounded by a spherical surface $S(V)$ of radius $R\gg \pi c/\omega$,
\begin{multline} \label{eq:intS}
  \oiint\limits_{S(V)} \left( {\bf E}^{(2\omega)} \times {\bf H}^{(2\omega)}_1  - {\bf E}^{(2\omega)}_1 \times {\bf H}^{(2\omega)}  \right)\cdot  {\VF{\dif{\hat{S}}}} \\= \dfrac{4\pi}{c} \iiint\limits_V {\bf j}^{({2\omega})} {\bf E}^{({2\omega})}_1 \dif V  \:.
 \end{multline}
As an auxiliary solution $ \left\{ {\bf{E}}^{\left( {2 \omega} \right) }_1,{\bf{H}}^{\left( {2 \omega} \right)}_1 \right\}$, we choose the electromagnetic field, which at $r>a$ constitutes the incident and reflected multipolar electric or magnetic mode. % wave
It acquires the following form for the electric multipolar mode:
\begin{equation}  \label{eq:AuxESph}
\begin{aligned}
{\bf{H}}^{\left( {2 \omega} \right)}_1 (r>a) & =  \left( h^{(2)}_l(2k_0r) +\eta^E_l h^{(1)}_l(2k_0r) \right){\bf{X}}_{l,m} ,  \\
{\bf{H}}^{\left( {2 \omega} \right)}_1 (r<a) & =  t^E_l j_l(2k_0 \sqrt{\eps(2\omega)} r ) {\bf{X}}_{l,m} \:, \\
{\bf{E}}^{\left( {2 \omega} \right)}_1 (r>a) & =  \dfrac{i}{2k_0} \nabla \times {\bf{H}}^{\left( {2 \omega} \right)}_1 ,  \\
{\bf{E}}^{\left( {2 \omega} \right)}_1 (r<a) & = \dfrac{i}{2k_0{\eps(2\omega)}} \nabla \times {\bf{H}}^{\left( {2 \omega} \right)}_1  \:,
\end{aligned}
\end{equation}
and for the magnetic multipolar mode:
\begin{equation} \label{eq:AuxMSph}
\begin{aligned}
{\bf{E}}^{\left( {2 \omega} \right)}_1 (r>a) & =  \left( h^{(2)}_l(2k_0r) +\eta^M_l h^{(1)}_l(2k_0r) \right){\bf{X}}_{l,m} ,  \\
{\bf{E}}^{\left( {2 \omega} \right)}_1 (r<a) & =  t^M_l j_l(2k_0 \sqrt{\eps(2\omega)} r ) {\bf{X}}_{l,m} \:, \\
{\bf{H}}^{\left( {2 \omega} \right)}_1  & =  -\dfrac{i}{2k_0} \nabla \times {\bf{E}}^{\left( {2 \omega} \right)}_1   \:.
\end{aligned}
\end{equation}
Here the Hankel function of the second kind $h^{(2)}_l(2k_0r)$ corresponds to the incident spherical % multipolar %(spherical?)
wave, while the Hankel function of the first kind $h^{(2)}_l(2k_0r)$ describes the reflected out-going mode. The spherical Bessel function $j_l(2k_0 \sqrt{\eps(2\omega)} r )$  describes the auxiliary field inside the nanoparticle. Reflection (transmission) coefficients $\eta^{E,M}_l$ ($t^{E,M}_l$) are derived from the condition of the continuity for the tangential component of the electric field and magnetic fields at the interface $r=a$:
\begin{widetext}
\begin{equation}  \label{eq:coeff}
\begin{aligned}
& \eta^M_l =  \left. -\dfrac{h^{(2)}_l(2k_0r) \partial_{r}\left[ r j_l(2k_0\sqrt{\eps(2\omega)}r)\right] - j_l(2k_0\sqrt{\eps(2\omega)}r) \partial_{r}\left[ r h^{(2)}_l(2k_0r)\right]}{h^{(1)}_l(2k_0r) \partial_{r}\left[ r j_l(2k_0\sqrt{\eps(2\omega)}r)\right] - j_l(2k_0\sqrt{\eps(2\omega)}r) \partial_{r}\left[ r h^{(1)}_l(2k_0r)\right]} \right|_{r=a},  \\
& t^M_l =  \left. \dfrac{i }{k_0 a} \left( j_l(2k_0\sqrt{\eps(2\omega)}r)  \partial_{r}\left[ r h^{(1)}_l(2k_0r)\right] -  h^{(1)}_l(2k_0 r)  \partial_{r}\left[ r j_l(2k_0\sqrt{\eps(2\omega)}r)\right] \right)^{-1}\right|_{r=a}\:, \\
& \eta^E_l = \left.  -\dfrac{( \eps(2\omega) )^{-1} h^{(2)}_l(2k_0r) \partial_{r}\left[ r j_l(2k_0\sqrt{\eps(2\omega)}r)\right] - j_l(2k_0\sqrt{\eps(2\omega)}r) \partial_{r}\left[ r h^{(2)}_l(2k_0r)\right]}{( \eps(2\omega) )^{-1} h^{(1)}_l(2k_0r)  \partial_{r}\left[ r j_l(2k_0\sqrt{\eps(2\omega)}r)\right] - j_l(2k_0\sqrt{\eps(2\omega)}r) \partial_{r}\left[ r h^{(1)}_l(2k_0r)\right]}\right|_{r=a}\: ,  \\
& t^E_l =  \left. \dfrac{i }{k_0 a} \left( j_l(2k_0\sqrt{\eps(2\omega)}r)  \partial_{r}\left[ r h^{(1)}_l(2k_0r)\right] - ( \eps(2\omega) )^{-1}  h^{(1)}_l(2k_0 r)  \partial_{r}\left[ r j_l(2k_0\sqrt{\eps(2\omega)}r)\right] \right)^{-1} \right|_{r=a} \:,
\end{aligned}
\end{equation}
\end{widetext}
Note that $ |\eta^{E,M}_l|= 1$ in Eq.~\eqref{eq:coeff}.

Substituting Eqs.~\eqref{eq:AuxESph} and~\eqref{eq:AuxMSph} into Eq.~\eqref{eq:intS}, with account for the expansion~\eqref{eq:SHMField} and the orthogonality condition for spherical harmonics, we get the excitation coefficients of the multipolar modes at the SH frequency:
\begin{widetext}
\begin{equation}  \label{eq:coeff2}
\begin{aligned}
& q^E_1 =  \dfrac{4\pi ik_0}{c} \dfrac{t^E_2}{E_0^2 (A^M_{1,1})^2 \eps (2\omega)} \iiint\limits_{r<a}  \dif V   {\bf j}^{({2\omega})} \cdot \nabla \times \left\{  j_2(2k_0\sqrt{\eps(2\omega)}r) {\bf{X}}_{2,0} (\theta_1) \right\},  \\
& q^E_2 =  \dfrac{4 \pi i k_0}{c} \dfrac{t^E_2}{E_0^2 (A^E_{1,1})^2 \eps (2\omega)} \iiint\limits_{r<a}  \dif V   {\bf j}^{({2\omega})} \cdot \nabla \times \left\{  j_2(2k_0\sqrt{\eps(2\omega)}r) {\bf{X}}_{2,0} (\theta_2) \right\}\:, \\
& d^E =  \dfrac{4\pi i k_0}{c} \dfrac{t^E_1}{E_0^2 A^E_{1,1} A^M_{1,1} \eps (2\omega)} \iiint\limits_{r<a}  \dif V   {\bf j}^{({2\omega})} \cdot \nabla \times \left\{  j_1(2k_0\sqrt{\eps(2\omega)}r) {\bf{X}}_{1,0} (\theta) \right\} ,  \\
& q^M=  \dfrac{8 \pi k_0^2}{c} \dfrac{t^M_2}{E_0^2 A^E_{1,1} A^M_{1,1} } \iiint\limits_{r<a}  \dif V   {\bf j}^{({2\omega})} \cdot   j_2(2k_0\sqrt{\eps(2\omega)}r) \left\{ {\bf{X}}_{2,1} (\theta_1,\phi_1)  -  {\bf{X}}_{2,-1} (\theta_1,\phi_1)  \right\} \:,
\end{aligned}
\end{equation}
\end{widetext}
Coefficients $q^E_1$, $q^E_2$, $d^E$ and $q^M$ are in different ways related to the phenomenological parameters $\gamma$, $\delta'$, $\chi^{\left( {2} \right)}_{ \bot  \bot  \bot }$, $\chi^{\left( {2} \right)}_{\bot \parallel \parallel} $, $\chi^{\left( {2} \right)}_{\parallel  \bot \parallel }$. In particular, calculating the integrals in the right-hand side of Eqs.~\eqref{eq:coeff2}, one can show that
\begin{equation}
\begin{aligned}
q^E_1 &  =  a_1  \gamma + a_2 \delta' + a_3 \chi^{\left( {2} \right)}_{\bot \parallel \parallel}\:,  \\
 q^E_2 & = b_1 \gamma + b_2 \delta' + b_3  \chi^{\left( {2} \right)}_{ \bot  \bot  \bot } +b_4 \chi^{\left( {2} \right)}_{\bot \parallel \parallel}  + b_5 \chi^{\left( {2} \right)}_{\parallel  \bot \parallel } \:, \\
d^E &  =  c_1 \gamma + c_2 \delta' + c_3   \chi^{\left( {2} \right)}_{\bot \parallel \parallel}\:,  \\
q^M & =  f \chi^{\left( {2} \right)}_{\parallel  \bot \parallel } \:,
\end{aligned}
\end{equation}
where coefficients $a_i$, $b_k$, $c_j$ and $f$ depend on the frequency, the particle size and the dielectric permittivity.
Remarkably, the magnetic quadrupolar component $q^M$ depends only on one parameter $\chi^{\left( {2} \right)}_{\parallel  \bot \parallel }$, and, hence, it vanishes at  $\chi^{\left( {2} \right)}_{\parallel  \bot \parallel }=0$.

At long distances from the particle, where $2 k_0 r\gg 1$, the electric field emitted at the second harmonic takes the following form:
\begin{widetext}
\begin{multline} \label{eq:longdistfield}
{\bf{E}}^{\left( {2\omega} \right) } ( {\bf r} ) \approx  {E_0}^2  \left[
- \dfrac{1}{4} \sqrt{\dfrac{15}{2\pi}} (A^M_{1,1})^2 q^E_1  \sin 2\theta_1  \bm{\hat{\theta}}_1  \right.  \left. - \dfrac{1}{4} \sqrt{\dfrac{15}{2\pi}} (A^E_{1,1})^2 q^E_2   \sin 2\theta_2  \bm{\hat{\theta}}_2+  \right.  \\ \left. A^E_{1,1} A^M_{1,1} \left\{ - \dfrac{i}{2} \sqrt{\dfrac{3}{2\pi}} d^E \sin \theta \bm{\hat{\theta}} - \sqrt{\dfrac{5}{4\pi}} (\cos{2\theta_1} \sin \phi_1 \bm{\hat{\phi}}_1 - \cos{\theta_1} \cos \phi_1 \bm{\hat{\theta}}_1 ) q^M \right\} \right] \dfrac{\exp{(2ik_0 r)}}{2 k_0 r}
\end{multline}
\end{widetext}
Here, $\bm{\hat{\theta}}_1$, $\bm{\hat{\theta}}_2$, $\bm{\hat{\theta}}$ are the unit vectors directed along the increasing polar variables in spherical coordinate systems associated  with the $y$, $x$ and $z$ axes, respectively, and $\bm{\hat{\phi}}_1$ is the corresponding azimuthal unit vector. The study of the SH radiation pattern in different cross sections may assist in estimating the relative values of the nonlinear
phenomenological parameters, $\gamma$, $\delta'$, $\chi^{\left( {2} \right)}_{ \bot  \bot  \bot }$, $\chi^{\left( {2} \right)}_{\bot \parallel \parallel} $, $\chi^{\left( {2} \right)}_{\parallel  \bot \parallel }$, of the quadratic nonlinearity of the
dielectric material the nanoparticle is made of~\cite{Wiecha2015,Wiecha2016}. 

\begin{figure} [t]
\includegraphics[width=0.97\columnwidth]{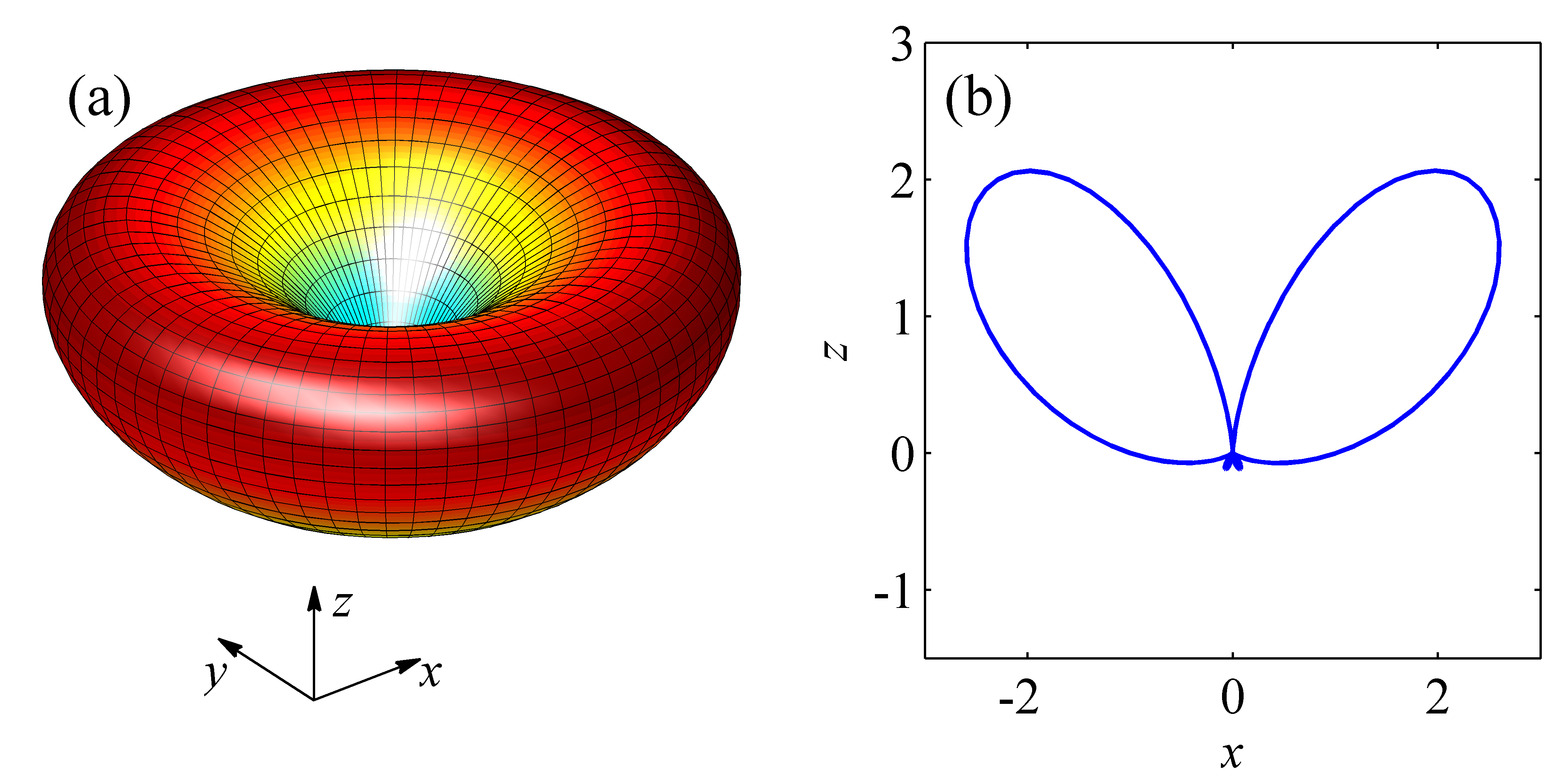} % image096.jpg
\caption{(Color online) Radiation pattern (a) of the second-harmonic radiation and its cross section (b) by the $xz$  plane  for
 $(A_{1,1}^M )^2 q^E_1 \approx (A_{1,1}^M )^2 q^E_2 \approx - i A_{1,1}^E A_{1,1}^M d_E$, $|d_E |\gg|q_M |$.
}  \label{fig:fig2}
\end{figure}

Note, being based on Lorentz lemma, our approach can be regarded as a more practical alternative to the analysis suggested in works~\cite{Biris2010,Gonella2011,Capretti2013}. In particular, the contribution of the truly volume separable $\delta'$ polarization source is found here not attracting a more involved treatment based on Green's function formalism.
While usually disregarded for metal nanostructures, the $\delta'$ source dependent on spatial derivatives of the fields inside in the nanoparticle is not necessarily negligible in dielectric nanoparticles. For the surface nonlinearity, we additionally check that the excitation coefficients for the dominating SH multipoles obtained with Eqs.~\eqref{eq:coeff2} in fact coincide with those recovered with the use of the nonlinear Mie theory for SHG from a spherical centrosymmetric nanoparticle~\cite{Dadap2004,Gonella2011}. For comparison, we employed the formulas given in Supplemental Material of Ref.~\cite{Gonella2011}.

The interference of the nonlinearly generated multipoles could be employed for engineering the radiation directionality.
Figure~\ref{fig:fig2} clearly demonstrates the possibility to implement a nonlinear antenna which
generates the second-harmonic light directionally. This directivity is achieved due to the excitation of the mutually perpendicular electric quadrupoles and a dipole, which are oriented along the $x$, $y$ and $z$ axes. The  exemplary radiation pattern in Fig.~\ref{fig:linscat} is plotted, assuming that the contribution of the magnetic quadrupole is small and the amplitudes of the ED and EQ modes in the SH radiation field are of the same order of magnitude.

\begin{figure} [t]
\includegraphics[width=1\columnwidth]{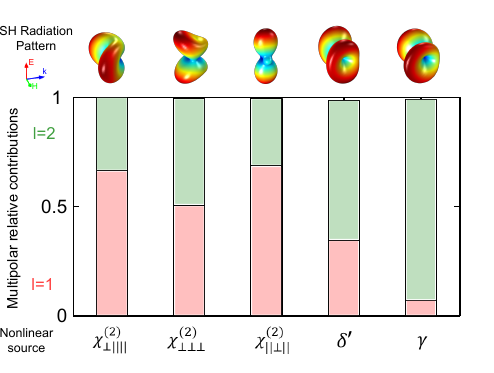}
\caption{(Color online) Numerically calculated multipolar decomposition of SHG from a spherical silicon nanoparticle of radius $a=145$~nm under plane-wave excitation at the pump wavelength $\lambda_0=1050$~nm and corresponding SH emission patterns for different nonlinear sources. Stacked bars visualize contributions of the dominating dipolar ($l=1$) and quadrupolar ($l=2$) orders normalized to the total SH radiated power. }  \label{fig:FigPmles}
\end{figure}

Our analytical considerations are confirmed by full-wave numerical modeling performed with finite-element-solver COMSOL Multiphysics, following the procedure described in Refs.~\cite{Smirnova:2016:ACS-Ph,Shorokhov2016,CamachoMorales2016,Wang2017,Kruk2017}. These simulations allow for solving the full scattering problem at the SH frequency using the induced nonlinear polarization within the undepleted pump approximation in the presence of the dielectric environment. Then the multipolar amplitude coefficients dependent on the geometry and a refractive-index contrast are retrieved~\cite{Grahn2012}. Because the values of the phenomenological nonlinear coefficients for silicon (Si) are yet not well established and it is the subject of discussion up to now, we examined different terms in nonlinear sources~\eqref{eq:PNL} separately, as if they acted independently, for silicon nanoparticles exhibiting overlapped MD and ED resonances, varying the radius in the range as shown in Fig.~\ref{fig:linscat}. In agreement with our theoretical model, for smaller radii $a \lesssim 170$~nm the leading contributions to the radiated SH field stem from the dipolar and quadrupolar modes we distinguished, as exemplified in Fig.~\ref{fig:FigPmles} for the nanoparticle of radius $a =145$~nm. Fig.~\ref{fig:FigPmles} evidences that the multipolar expansion of SHG up to the order $l = 2$ well approximates the total radiated power, while the higher-order corrrections $l \ge 3$ appear small. In this regime SHG process is essentially governed by two dipolar modes excited at the fundamental wavlength, because of their resonant character [Fig.~\ref{fig:linscat}], which distinguishes the case under study e.g. from the described in the literature SHG by small nonresonant nanoparticles and low-index-contrast polysterene nanoparticles in water~\cite{Dadap2004,Shan2006,Wunderlich2011}. COMSOL results additionally confirm that when defining the SH nonlinear source through the bulk and surface nonlinear polarizations, one may, to a high degree of accuracy, restrict oneself to taking into account electric and magnetic dipolar modes only. This is reasonably explained by sufficiently high quality factors of the dipolar resonances exhibiting by the high-index nanoparticles of the corresponding sizes. 

\begin{figure} [h]
\includegraphics[width=0.95\columnwidth]{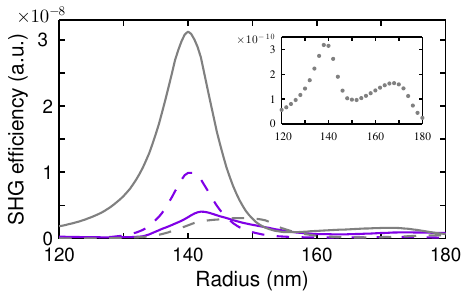}
\caption{ (Color online) Second-harmonic conversion efficiency calculated numerically from a spherical silicon nanoparticle at the incident intensity $I_0=1$ GW/cm$^2$ and laser wavelength $\lambda_0=1050$~nm.
The dependences of different source contributions on
the nanoparticle radius were computed independently: $\chi^{\left( {2} \right)}_{\bot \parallel \parallel}$ (solid gray), $\chi^{\left( {2} \right)}_{\parallel  \bot \parallel }$ (dashed gray), $\gamma$ (dashed purple), $\delta'$ (solid purple).
The inset shows $\chi^{\left( {2} \right)}_{ \bot  \bot  \bot }$ SHG efficiency (dotted gray).}   \label{fig:FigPSH}
\end{figure}

We expect the total conversion efficiency to be dispersive and size-dependent. It is strongly affected by hierarchy of Mie resonances and modal overlaps, as was shown experimentally for Mie-resonant nanoparticles in the recent works~\cite{CamachoMorales2016, Makarov2017,Kruk2017}. Direct numerical simulations performed with COMSOL reveal that with increasing the nanoparticle's size (closer to $a=200$ nm), the higher orders (up to $l=4$) show up in the multipolar expansion of the SH field. Based on the data and discussions in Refs.~\cite{Palik,Falasconi2001,Mochn2003, Wiecha2016}, we approximately estimate the SHG efficiency and bulk and surface relative contributions % ratio
for a silicon nanoparticle under
the plane-wave illumination [Fig.~\ref{fig:FigPSH}]. For calculations, we take $\chi^{\left( {2} \right)}_{ \bot  \bot  \bot } = 65 \times 10^{-19}$ m$^2$/V,
and set the other nonlinearity parameters roughly of the same order of magnitude, $\chi^{\left( {2} \right)}_{\bot \parallel \parallel} = \chi^{\left( {2} \right)}_{\parallel  \bot \parallel } =\gamma =3.5\times 10^{-19}$ m$^2$/V. Using the polarizability model~\cite{Mochn2003}, we estimate $\delta'\approx \gamma \left( \eps(\omega) -4\eps(2\omega) + 3 \right) /\left(\eps(2\omega) -1 \right)$. The pronounced enhancement in the second-harmonic (SH) scattering occurs near MD resonance at the pump wavelength. The dominant peaks are exhibited by $\chi^{\left( {2} \right)}_{\bot \parallel \parallel}$ surface and $\gamma$, $\delta'$ bulk sources, which we analyze in detail in Sec.~III.

However, the volume response can be attributed exclusively to the separable bulk $\delta'$ term.
The mentioned above quasisurface character of the bulk $\gamma$ term~\cite{Capretti2013,Wiecha2016} can be inferred from Eqs.~\eqref{eq:coeff2}. Since $q^M$ depends only on $\chi^{\left( {2} \right)}_{\parallel  \bot \parallel }$, we inspect the amplitudes of electric modes $q^E_{1,2}$ and $d^E$. For the $\gamma$ source, they can be transformed to the surface integrals as follows
\begin{multline}
q^E_{1,2},\:d^E \propto \gamma \iiint\limits_{r<a} \dif V  \nabla \left({\bf{E}}^{\left( {\omega} \right) }_{\text{in}}\right)^2  \cdot \left[ \nabla \times \left\{ j_{2(1)} {\bf{X}}_{2(1),0} \right\} \right]\\
=\gamma \iiint\limits_{r<a} \dif V  \nabla \cdot \left[ \nabla \times \left\{  \left( {\bf{E}}^{\left( {\omega} \right) }_{\text{in}} \right)^2 j_{2(1)} {\bf{X}}_{2(1),0}  \right\} \right]\\
= \gamma \oiint\limits_{r=a} {\VF{\dif{\hat{S}}}}  \left( {\bf{E}}^{\left( {\omega} \right) }_{\text{in}}\right)^2  \left[ \nabla \times j_{2(1)} {\bf{X}}_{2(1),0}  \right]\:,
\end{multline}
justifying that the bulk $\gamma$ term contributes to the effective surface response.

\section{Second-harmonic generation driven by magnetic dipole mode}

In this section, we consider in more detail and derive an analytical solution for SHG from a high-index dielectric particle driven by the MD mode. This particular case well describes the pronounced magnetic dipole resonance~\cite{Smirnova:2016:ACS-Ph}. Alternatively, it may be realized in experiment by irradiating the nanoparticle with the azimuthally-polarized beam whose structure imitates the MD mode
polarization distribution. In this instance, solution can be obtained from the analysis developed in Sec.~II  by setting $A^E_{1,1}=0$. However, for the sake of methodological clarity, here  we take a different way and solve this basic nonlinear problem not attracting the Lorentz lemma but following the approach outlined in Ref.~\cite{Smirnova:2016:ACS-Ph}, where the third-harmonic generation by resonant silicon nanoparticles was described.

We employ a single-mode approximation and assume the fields inside the nanoparticle at $r<a$ % ($\sim e^{-i\omega t}$)
are given by MD mode profile as follows
\begin{equation}
\label{eq:Field_in_VSH_mdS}
\begin{split}
& {\bf{E}}^{\left( {\omega} \right)}_{\text{in}} \! \approx \!  A(\omega) j_1(k(\omega) r)\left\{ \bm {\hat{\theta}} \cos \phi - \sin \phi \cos \theta \bm{\hat{\phi}} \right\}\:,
\\
& {\bf{H}}^{\left( {\omega} \right) }_{\text{in}} \! \approx \!  \dfrac{A(\omega)}{ik_0} \nabla \times \left[ j_1(k(\omega) r)\left\{ \bm {\hat{\theta}} \cos \phi - \sin \phi \cos \theta \bm{\hat{\phi}} \right\} \right],
\end{split}
\end{equation}
where $A(\omega)=E_0 A^{M}_{1,1} \dfrac{1}{2} \sqrt{\dfrac{3}{\pi}} $. We rewrite expressions~\eqref{eq:Field_in_VSH_mdS} in the spherical coordinate system associated with $y$ axis co-directed with the induced magnetic dipole moment:
{\small{
\begin{equation} \label{eq:Field_in_VSH_mdS2}
\begin{split}
& {\bf{E}}^{\left( {\omega} \right) }_{\text{in}}\! \approx \!   A(\omega) j_1(k(\omega) r) \sin{\theta}_1 \bm{\hat{\phi}}_1 \:,
\\
 & {\bf{H}}^{\left( {\omega} \right) }_{\text{in}} \! \approx \!   % \dfrac{A(\omega)}{ik_0} \nabla \times \left\{ j_1(k(\omega) r) \sin{\theta}_1  \bm{\hat{\phi}_1} \right\} =
 \dfrac{A(\omega)}{ik_0} \left\{ \bm{\hat{r}} \dfrac{2\cos {\theta}_1 }{r}  j_1(k(\omega) r) - \bm{\hat{\theta}}_1  \dfrac{\sin{\theta}_1 }{r} \partial_r \left( r j_1 (k(\omega) r) \right) \right\}.
\end{split}
\end{equation}
}}
Substituting the fields~\eqref{eq:Field_in_VSH_mdS2} into Eqs.~\eqref{eq:PNL}, the  nonlinear polarizations are recast to the surface source caused solely by the tensor component $\chi _{\bot \parallel \parallel} $,
\begin{multline} \label{eq:surf}
{\bf{P}}_{\text{surf}}^{^{\left( {2\omega } \right)}} =
%\bm{\hat{r}} \chi _{\bot \parallel \parallel} \bm{E}_{\tau} \bm{E}_{\tau} \delta(r-a-0) =
\bm{\hat{r}} \chi _{\bot \parallel \parallel} ({\bf {E}}^{\left( {\omega} \right) }_{\text{in}})^2\delta(r-a+0)
\\ = \bm{\hat{r}} A^2(\omega) \chi _{\bot \parallel \parallel} j_1^2(k (\omega) r) \sin^2{\theta}_1\delta(r-a+0) \\
=\bm{\hat{r}} {{P}}_{\text{s}}^{^{\left( {2\omega } \right)}} \delta(r-a+0) ,
\end{multline}
and the bulk source consisting of two regrouped contributions
{\small{
\begin{multline} \label{eq:bulk}
{\bf{P}}_{\text{bulk}}^{^{\left( {2\omega } \right)}} \!=\!
 \left\{ \gamma \nabla \left( {{{\bf{E}}^{\left( \omega  \right)}} \cdot {{\bf{E}}^{\left( \omega  \right)}}} \right)\! +\!
  \; \delta' \left( {{{\bf{E}}^{\left( \omega  \right)}} \cdot \nabla } \right){{\bf{E}}^{\left( \omega  \right)}} \right\} \Pi(a-r-0) \\
  =\left\{ (\gamma + \delta'/2) \nabla ({\bf{E}}^{\left( {\omega} \right) }_{\text{in}})^2 - \delta'\left[ {\bf{E}}^{\left( {\omega} \right) }_{\text{in}} \times i k_0 {\bf H}^{\left( {\omega} \right) }_{\text{in}} \right] \right\}  \Pi(a-r-0) \\
  = {\bf{P}}_{\text{bulk1}}^{^{\left( {2\omega } \right)}} +  {\bf{P}}_{\text{bulk2}}^{^{\left( {2\omega } \right)}} \:.
\end{multline}
}}
% which we decompose into two contributions, respectively.
\begin{comment}
{\small{
\begin{subequations}
\begin{align}
&  {\bf{P}}_{\text{bulk}1}^{^{\left( {2\omega } \right)}} =A^2(\omega)  (\gamma + \delta'/2) \nabla \left(  j^2 _1 (k (\omega) r) \sin^2 {\theta}_1 \right) \Pi(a-r-0) \:, \\
\begin{split}
 & {\bf{P}}_{\text{bulk}2}^{^{\left( {2\omega } \right)}} = - \delta' A(\omega)^2 \left\{ \sin^2 \theta' j_1(k_2 r) \dfrac{1}{r} \dfrac{\partial}{\partial r} r j_1(k_2 r) \bm{\hat{r}} \right. \\ & \left.  + \dfrac{1}{r} j_1^2(k_2 r) \sin 2 \theta' \bm{\hat{\theta}_1} \right\}
\text{ ,}
\end{split}
\end{align}
\end{subequations}
}}
\end{comment}
For clarity, we consider response of the structure driven by
the % these three % induced
nonlinear sources ${\bf{P}}_{\text{surf}}^{^{\left( {2\omega } \right)}}$, ${\bf{P}}_{\text{bulk1}}^{^{\left( {2\omega } \right)}}$ and ${\bf{P}}_{\text{bulk2}}^{^{\left( {2\omega } \right)}}$ sequentially.

The normal surface polarization~\eqref{eq:surf} in the driven Maxwell's equations is equivalent to the dipole layer. Alternatively, in electrodynamic equations it may be formally replaced by the fictitious surface magnetic current whose density is defined by
\begin{equation}
{\bf {j}}^{(2\omega) M}_{\text{surf}1} = \dfrac{c}{ \eps(2\omega)}\left. [\nabla {{P}}_{\text{s}}^{^{\left( {2\omega } \right)}} \times \bm{\hat{r}} ]\right|_{r=a-0}\:.
\end{equation}
Thus,
% the surface source~\eqref{eq:surf} leads to the jump of
% undergoes a jump  experiences
the tangential $\theta_1$-component of the electric field at the spherical boundary $r=a$ undergoes a jump expressed through the derivative $\partial_{{\theta}_1} {{P}}_{\text{s}}^{^{\left( {2\omega } \right)}} $:
\begin{multline} \label{eq:bound1}
\left. E_{{\theta}_1}^{(2\omega)} \right|_{r=a+0} - \left. E_{{\theta}_1}^{(2\omega)}\right|_{r=a-0}  = -\dfrac{4\pi} {\left(\eps(2\omega)a \right)} \dfrac{\partial {{P}}_{\text{s}}^{^{\left( {2\omega } \right)}} }{\partial \theta_1}\\
=- 4\pi \left(\eps(2\omega)a \right)^{-1} \chi _{\bot \parallel \parallel} A^2(\omega) j_1^2(k(\omega)a)\sin 2\theta_1.
\end{multline}

Considering the term ${\bf{P}}_{\text{bulk1}}^{^{\left( {2\omega } \right)}}$, which is a gradient of the scalar function, we
% ${\bf{P}}_{\text{bulk}1}^{^{\left( {2\omega } \right)}}= (\gamma + \delta'/2) \nabla ({\bf E}^{\left( {\omega} \right) }_{\text{in}})^2 $
represent the electric field as a sum of the vortex and potential vector fields
\begin{equation} \label{eq:subst}
{\bf E}^{(2\omega)} = {\bf E}^{(2\omega)}_{\text{v}}- \dfrac{4\pi}{\eps(2\omega)} \nabla ({\bf E}^{\left( {\omega} \right) }_{\text{in}})^2 (\gamma + \dfrac{\delta'}{2})   \begin{cases}
1, & r<a\\
0, & r>a\
\end{cases}
\end{equation}
The vortex part $ {\bf E}_{\text{v}}^{(2\omega)}$ is, therefore, found by solving the Maxwell's equations with the substitution~\eqref{eq:subst} transformed to % reorganized
\begin{equation}
\begin{cases}
\nabla \times {\bf E}^{(2\omega)}_{\text{v}} = 2ik_0 {\bf H}^{(2\omega)}, & \\
\nabla \times {\bf H}^{(2\omega)} = -2ik_0 {\bf E}^{(2\omega)}_{\text{v}} \eps^{(2\omega)} (r)\:,
\end{cases}
\end{equation}
with the following boundary conditions at the nanoparticle surface
\begin{subequations} \label{eq:bound2}
\begin{align}
% & \left. E^{(2\omega)}_{\text{v}{r}} \right|_{r=a+0} =\eps(2\omega) \left. E_{{\text{v}}{r}}^{(2\omega)}\right|_{r=a-0}
& \left. H^{(2\omega)}_{{\phi}_1} \right|_{r=a+0} =\left. H_{{\phi}_1}^{(2\omega)}\right|_{r=a-0}, \label{eq:bound2a}  \\
& \left. E_{{\text{v}}{{\theta}_1}}^{(2\omega)} \right|_{r=a+0} - \left. E_{{\text{v}}{{\theta}_1}}^{(2\omega)}\right|_{r=a-0} = \left. \dfrac{4\pi}{c} (  {\bf j}^{(2\omega)M}_{\text{surf}2}({2\omega}) \bm{\hat{\phi}}_1)\right|_{r=a}\:, \label{eq:bound2b}
\end{align}
\end{subequations}
where, to account for the electric field discontinuity at the interface $r=a$, we have again introduced the surface magnetic current given by % at $r=a-0$ defined by
\begin{multline}
{\bf j}^{(2\omega)M}_{\text{surf}2}=-\dfrac{c} {\eps(2\omega)}(\gamma + \delta'/2)[\bm{\hat{r}} \times \nabla ({\bf E}^{\text{in}}_{(\omega)})^2 ]   = \\
- \dfrac{c}{\eps(2\omega)}A^2(\omega) (\gamma + \dfrac{\delta'}{2})\dfrac{1}{a} j_1^2(k(\omega)a)\sin 2\theta_1 {\bm{\hat{\phi}}}_1  \:.
\end{multline}
\begin{comment}
leading to the jump
\begin{equation}
\left. E_{{\theta'} {\sim}}(2\omega) \right|_{r=a+0} - \left. E_{{\theta'} {\sim}}(2\omega)\right|_{r=a-0} = \left. \dfrac{4\pi}{c} (  {\bf j}^{\text{m}}_{s}({2\omega}) \bm{\hat{\phi}'})\right|_{r=a} =
- 4\pi \dfrac{\gamma + \delta'/2}{\eps(2\omega)} A^2(\omega)\dfrac{1}{a} j_1^2(k(\omega)a)\sin{2\theta'} ,
\end{equation}
\end{comment}

Noticeably, the boundary conditions~\eqref{eq:bound1} and~\eqref{eq:bound2b} can be additively combined to
\begin{multline} \label{eq:boundcombined}
\left.  E_{{\text{v}}{{\theta}_1}}^{(2\omega)} \right|_{r=a+0} - \left. E_{{\text{v}}{{\theta}_1}}^{(2\omega)}\right|_{r=a-0}   = \\
-\dfrac{4\pi}{{a}{\eps(2\omega)}} \left( \chi _{\bot \parallel \parallel}  + \gamma +\delta'/2 \right) A^2(\omega)j_1^2(k(\omega)a)\sin{2\theta_1}\:.
\end{multline}

\begin{figure*} [t]
 	\includegraphics[width=14 cm]{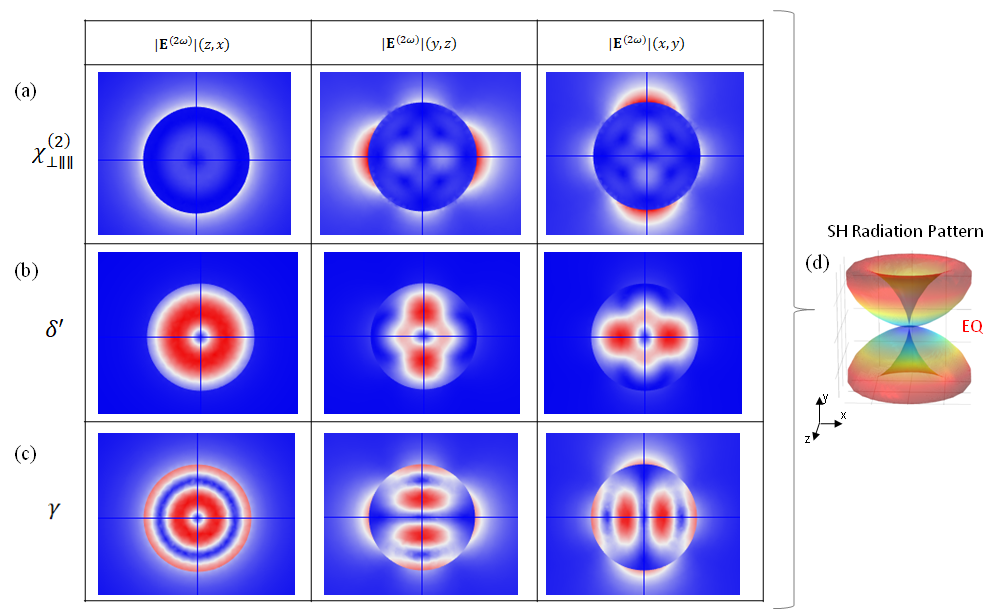}
 	\caption{(Color online) SHG in Si nanoparticle driven by MD mode. Simulated SH field distributions generated by different SH nonlinear sources % induced due to
    stemming from (a) $ \chi^{(2)}_{\bot \parallel \parallel} $, (b) $\delta'$, (c) $\gamma$. The nonlinear response is set to be driven by MD mode associated with $y$-polarized magnetic dipole moment at the fundamental frequency.
 		(d) The nanoparticle radiates SH light as EQ in all three cases.
 		Computed for the pump wavelength $\lambda_0=1050$~nm and nanoparticle radius $a=145$~nm.}  \label{fig:SHdrivenMD}
 \end{figure*}

With the second part of the bulk source, being nonzero only if $\delta'\ne 0$,
\begin{multline}
{\bf{P}}_{\text{bulk}2}^{^{\left( {2\omega } \right)}}= - \delta' A^2(\omega) \left\{ j_1(k(\omega)r) \dfrac{1}{r} \dfrac {\partial}{\partial r} r j_1(k(\omega) r) \sin^2{\theta_1} \bm{\hat{r}} \right.\\
\left. + \dfrac{1}{r} j_1^2(k(\omega) r) \sin 2\theta_1 {\bm{\hat{\theta}}}_1\right\} \Pi(a-r-0)\:,
\end{multline}
inside the particle at $r<a-0$ we solve the inhomogeneous wave equation:
\begin{equation}
\nabla \times \nabla \times {\bf H}^{(2\omega)} - 4k_0^2 \eps(2\omega)  {\bf H}^{(2\omega)} = 8 \pi ik_0 \nabla \times
{\bf{P}}_{\text{bulk}2}^{^{\left( {2\omega } \right)}}
\end{equation}
The solution is sought in the form $ {\bf H}^{(2\omega)} = H(r) \sin 2\theta_1 {\bm{ \hat{\phi}}}_1$, consistent with the angular structure of the source. % dependence
Remarkably, this corresponds to the electric quadrupole SH radiation in the far field.

Thereby, for the radial function $H(r)$ at $r<a$ we have equation
\begin{equation} \label{eq:radint}
\dfrac{d^2 H}{d r^2} + \dfrac{2}{r} \dfrac{dH}{dr}  -  \dfrac{6}{r^2} H + 4 k_0^2 \eps(2\omega)H \! = \! f (r)\:,
\end{equation}
with the source function $f(r)$ in the right-hand side:
\begin{multline}
f(r) = - 4\pi i k_0 \delta' A^2(\omega) \left\{ r  \dfrac{\partial}{\partial r} \left( \dfrac{j_1^2 (k(\omega)r) } {r^2} \right) \Pi(a-r-0) \right. \\ \left. - 2 \dfrac{j^2_1(k(\omega)r)}{r} \delta(r-a+0) \right\}\:.
\end{multline}
Solution of inhomogeneous second-order differential Eq.~\eqref{eq:radint} is
then found using Wronskian
\begin{multline}
H(r<a) =
 C_1 j_2(2k_0\sqrt{\eps(2\omega)} r)  % + 0 \cdot y_l( {K}_{2} r)
 + 2 k_0\sqrt{\eps(2\omega)}  \\
 \cdot \left( y_2(2k_0\sqrt{\eps(2\omega)} r) \int_{0}^{r} dr' r'^2 {f}(r') j_2(2k_0\sqrt{\eps(2\omega)} r') \right.\\
 \left. - j_2( 2k_0\sqrt{\eps(2\omega)} r) \int_{a}^{r} dr' r'^2 {f}(r') y_2( 2k_0\sqrt{\eps(2\omega)} r') \right)\:,
\end{multline}
where $y_2(2k_0\sqrt{\eps(2\omega)} r)$ is the spherical Neumann function.

Outside the nanoparticle at $r>a$ the magnetic field of the radiated SH electromagnetic quadrupolar wave is % outgoing spherical waves
\begin{equation}
{\bf H}^{(2\omega)} (r>a) = C_2 h^{(1)}_2(2k_0r) \sin 2\theta_1 {\bm{ \hat{\phi}}}_1\:.
\end{equation}
The efficiency of the SH quadrupolar radiation is determined by the coefficient $C_2$.

As follows from
conditions~\eqref{eq:bound2a} and ~\eqref{eq:boundcombined}, at the boundary $r=a$ the magnetic field $H^{(2\omega)}_{{\phi}_1}$ is continuous, while the $\theta_1$-component of the electric field experiences a jump caused by the fictitious surface magnetic current.
Matching these boundary conditions, we find the coefficient $C_2$ to be of the following form
\begin{multline} \label{eq:C2}
C_2 = - \dfrac{8 \pi k_0^2 a}{\eps(2\omega)} t^E_2 A^2(\omega) \left\{ \left(  \chi^{(2)}_{\bot \parallel \parallel}  + \gamma + \dfrac{3}{2}\delta' \right) \right. \\ \left. \cdot j^2_1(k(\omega)a) j_2 (2k_0 \sqrt{\eps(2\omega)}a) +  \dfrac{\delta'}{a} \int_{0}^{a} dr' \biggl( j^2_1(k(\omega)r')  \right. \\ \left.  - \dfrac{r'}{2} \dfrac{\partial}{\partial r'} j^2_1(k(\omega)r') \biggr) j_2 (2k_0 \sqrt{\eps(2\omega)}r')    \right\} \:.
\end{multline}
Substituting nonlinear sources~\eqref{eq:surf} and~\eqref{eq:bulk} into Eqs.~\eqref{eq:coeff2} and getting $q^E_1$, it can be seen that the amplitude of the electric quadrupolar mode given by
\begin{equation}
C_2 = E_0^2 (A^M_{1,1})^2 \dfrac{i}{4} \sqrt{{\dfrac{15}{2\pi}}} q^E_1\:
\end{equation}
is consistent with Eq.~\eqref{eq:C2}.
Thus, both the methods, based on (i) the Lorentz lemma [Sec.~II] and (ii) direct calculations of SH fields  [Sec.~III], yield the same result. However, in more involved situations, when SHG is governed by several multipoles excited at the fundamental frequency, approach (i) enables an easier way to recover analytical expressions for coefficients of multipolar expansion of nonlinear scattering.

Figure~\ref{fig:SHdrivenMD}(a,b,c) shows numerically calculated SH field near-field profiles generated by different nonzero source polarizations, associated with $ \chi^{(2)}_{\bot \parallel \parallel} $, $\delta'$, $\gamma$, for the case of pure MD mode excitation at the fundamental frequency. The $\chi^{(2)}_{\parallel \bot  \parallel} $ and $ \chi^{(2)}_{\bot \bot  \bot} $ SH sources vanish, given the absence of the electric field component normal to the surface. The total powers radiated by the nonzero sources relate in proportions consistent with Eq.~\eqref{eq:C2}. In agreement with our analytical results, in all three cases the simulated far-field manifests EQ structure, as depicted in Fig.~\ref{fig:SHdrivenMD}(d).

\section{Concluding remarks}

We have developed the theoretical model of the second-harmonic generation from high-index dielectric nanoparticles made of centrosymmetric materials (with a focus on silicon) excited by laser radiation in the frequency range covering the magnetic and electric dipolar Mie 
resonances at the fundamental frequency. We have shown that the multipolar decomposition of the generated second-harmonic field is dominated by the dipolar and quadrupolar modes. With the adjusted parameters, interference of these modes can ensure a good directivity of the SHG radiation.

We specifically focused on the magnetic dipole resonance inherent to high-permittivity dielectric nanoparticles and its influence on the  nonlinear scattering. It should be emphasized that magnetic modes bring new physics to simple dielectric geometries~\cite{kuznetsov2016science, Smirnova2016, Smirnova:2016:ACS-Ph, Kruk2017}, that differs substantially from the fundamentals of nonlinear nanoplasmonics largely appealing to the electric dipole resonances and electric modes, associated with surface plasmons~\cite{Dadap1999,Dadap2004,Gonella2011,Butet2012,Kauranen2012,Capretti2013,Smirnova2014,Butet2015}.
In particular, multipolar nature of nonlinear scattering is concerned.
As was established, both theoretically and experimentally, for the Rayleigh limit of SHG from a spherical metal nanoparticle under $x$-polarized plane-wave illumination, $z$-aligned $d^E$ electric dipole and $x$-axially symmetric $q^E_2$ electric quadrupole provide leading contributions to SH radiation, with zero SH signal in the forward direction.
By contrast, the excitation of magnetic dipole mode in dielectric nanoparticles may lead to generation of magnetic multipoles~\cite{Smirnova:2016:ACS-Ph,Kruk2017}. For instance, a silicon nanoparticle with cubic bulk
nonlinearity excited in the vicinity of magnetic dipole resonance produces third-harmonic radiation composed of magnetic dipole and octupole~\cite{Smirnova:2016:ACS-Ph}. The predominant generation of SH magnetic multipoles was also demonstrated experimentally in noncentrosymmetric AlGaAs nanodisks by tuning polarization of the optical pump~\cite{Kruk2017}. Here, we have shown that while the SH radiated field in the centrosymmetric nanoparticle driven by the magnetic dipole mode alone is solely constituted by the $q^E_1$ electric quadrupole spherical wave, the overlap of MD and ED modes under plane-wave excitation enriches the multipolar composition and brings magnetic quadrupole $q^M$ component.
The distinctive feature attributed to the magnetic dipole mode excitation is that the axis of the generated SH electric quadrupole $q^E_1$ is aligned with the magnetic moment at the pump wavelength, as illustrated in Fig.~\ref{fig:fig1}.

We believe our approach based on the Lorentz lemma is of a general nature, and, in combination with numerical calculations, it can be applied to describe the harmonic generation (such as SHG, THG) by Mie-resonant dielectric nanoparticles of an arbitrary shape, including those made of noncentrosymmetric materials, e.g. AlGaAs~\cite{CamachoMorales2016, Kruk2017} and BaTiO$_3$~\cite{Timpu2017,Ma2017}, which possess large volume quadratic susceptibility of a tensorial form. Our study and developed analytical approaches may be, therefore, instructive  for a design of efficient nonlinear all-dielectric nanoantennas with controllable radiation characteristics.

\section*{ACKNOWLEDGMENTS}

This work has been supported by the Russian Foundation for Basic Research (RFBR) (Grant 16-02-00547) and the Australian Research Council.

% \bibliographystyle{osajnl}
% \bibliography{SHG_nSilicon}

\end{document}